\documentclass[sn-mathphys,Numbered]{sn-jnl}% Math and Physical Sciences Reference Style
\usepackage[markup=underlined]{changes}
\definechangesauthor[color=red]{MS}
\definechangesauthor[color=blue]{xxx}
\usepackage{todonotes}
\setcommentmarkup{\todo[color={authorcolor!20},size=\scriptsize]{#3: #1}}
\usepackage{graphicx}%
\usepackage{multirow}%
\usepackage{amsmath,amssymb,amsfonts}%
\usepackage{amsthm}%
\usepackage{mathrsfs}%
\usepackage[title]{appendix}%
\usepackage{xcolor}%
\usepackage{textcomp}%
\usepackage{manyfoot}%
\usepackage{booktabs}%
\usepackage{algorithm}%
\usepackage{algorithmicx}%
\usepackage{algpseudocode}%
\usepackage{hyperref}       %

\usepackage[utf8]{inputenc} %
\usepackage[T1]{fontenc}    %

\usepackage{url}            %
\usepackage{booktabs}       %
\usepackage{nicefrac}       %
\usepackage{microtype}      %
\usepackage{xcolor}         %
\usepackage[export]{adjustbox}
\usepackage{wrapfig,amsthm,textcomp,amssymb}
\usepackage{colortbl}
\usepackage{pifont}%
\usepackage[capitalise,noabbrev,nameinlink]{cleveref}
\usepackage{adjustbox}

\usepackage{subfigure}
\usepackage[percent]{overpic}
\usepackage{lineno}

\usepackage{soul}
\usepackage{cancel}
\usepackage{graphicx}
\usepackage{listings}%
\definecolor{good}{HTML}{ffd2ad}
\definecolor{good}{HTML}{74c476}

\begin{document}

\title[Article Title]{Online Test-time Adaptation for Interatomic Potentials}


\author[1,2]{\fnm{Taoyong} \sur{Cui}}
\equalcont{These authors contributed equally to this work.}
\author[1,3,4]{\fnm{Chenyu} \sur{Tang}}
\equalcont{These authors contributed equally to this work.}
\author[1]{\fnm{Dongzhan} \sur{Zhou}}
\author[1]{\fnm{Yuqiang} \sur{Li}}
\author[5,6]{\fnm{Xingao} \sur{Gong}}
\author[1]{\fnm{Wanli} \sur{Ouyang}}
\author*[1]{\fnm{Mao} \sur{Su}}\email{sumao@pjlab.org.cn}
\author*[1]{\fnm{Shufei} \sur{Zhang}}\email{zhangshufei@pjlab.org.cn}

\affil[1]{\orgname{Shanghai Artificial Intelligence Laboratory}, \orgaddress{\city{Shanghai}, \postcode{200232}, \country{China}}}

\affil[2]{\orgname{Multimedia Laboratory, the University of Hong Kong}, \orgaddress{\city{Hongkong}, \postcode{999077}, \country{China}}}

\affil[3]{\orgdiv{Shenzhen International Graduate School}, \orgname{Tsinghua University}, \orgaddress{\city{Shenzhen}, \postcode{518055}, \country{China}}}

\affil[4]{\orgdiv{CAS Key Laboratory of Theoretical Physics}, \orgname{Institute of Theoretical Physics}, \orgname{Chinese Academy of Sciences}, \orgaddress{\city{Beijing}, \postcode{100190}, \country{China}}}

\affil[5]{\orgdiv{School of Physical Sciences}, \orgname{University of Chinese Academy of Sciences}, \orgaddress{\city{Beijing}, \postcode{100049}, \country{China}}}

\affil[6]{\orgdiv{Key Laboratory for Computational Physical Sciences (MOE), State Key Laboratory of Surface Physics, Department of Physics}, \orgname{Fudan University}, \orgaddress{\city{Shanghai}, \postcode{200433}, \country{China}}}

\affil[7]{\orgname{Shanghai  Qi Zhi Institute}, \orgaddress{\city{Shanghai}, \postcode{200232}, \country{China}}}


\abstract{Machine learning interatomic potentials (MLIPs) enable more efficient molecular dynamics (MD) simulations with ab initio accuracy, which have been used in various domains of physical science. However, distribution shift between training and test data causes deterioration of the test performance of MLIPs, and even leads to collapse of MD simulations. In this work, we propose an online Test-time Adaptation Interatomic Potential (TAIP) framework to improve the generalization on test data. Specifically, we design a dual-level self-supervised learning approach that leverages global structure and atomic local environment information to align the model with the test data. Extensive experiments demonstrate TAIP's capability to bridge the domain gap between training and test dataset without additional data. TAIP enhances the test performance on various benchmarks, from small molecule datasets to complex periodic molecular systems with various types of elements. Remarkably, it also enables stable MD simulations where the corresponding baseline models collapse.}


\maketitle

\section{Introduction}\label{sec1}

Molecular Dynamics (MD) simulation serves as a crucial technique across various disciplines including biology, chemistry, and material science \cite{hospital2015molecular, senftle2016reaxff, karplus1990molecular, yao_applying_2022}. MD simulations are typically based on interatomic potential functions that characterize the potential energy surface of the system, with atomic forces derived as the negative gradients of the potential energies. Subsequently, Newton's laws of motion are applied to simulate the dynamic trajectories of the atoms. In ab initio MD simulations \cite{car1985unified}, the energies and forces are accurately determined by solving the equations in quantum mechanics. However, the computational demands of ab initio MD limit its practicality in many scenarios. By learning from ab initio calculations, machine learning interatomic potentials (MLIPs) have been developed to achieve much more efficient MD simulations with ab initio-level accuracy \cite{butler_machine_2018, noe_machine_2020, unke_machine_2021}. 

Despite their successes, the crucial challenge of implementing MLIPs is the distribution shift between training and test data. When using MLIPs for MD simulations, the data for inference are atomic structures that are continuously generated during simulations based on the predicted forces, and the training set should encompass a wide range of atomic structures to guarantee the accuracy of predictions. However, in fields such as phase transition~\cite{zhang2021phase, niu2020ab}, catalysis~\cite{li2023situ, chen2024square}, and crystal growth~\cite{su2022exploring, zhang2024active}, the objective of MD simulations involves unveiling microscopic processes of rare events, in which the atomic structures are difficult to be captured especially in the initial training set. Consequently, a distribution shift between training and test datasets often occurs, which causes the degradation of test performance and leads to the emergence of unrealistic atomic structures, and finally the MD simulations collapse~\cite{fu2022forces}. Although strategies such as active learning~\cite{dp_gen, Uncertainty_driven, yuan2023active} and pretraining~\cite{zhang2022dpa, zhang2023dpa, denoise, gardner2024synthetic, cui2024geometry} have been developed to alleviate this challenge, they still struggle to explore unknown atomic structures and inevitably consume more computational resources. Therefore, a method that address the distribution shift without exploring additional atomic structures is desired.
%Therefore, a method that address the distribution shift without using additional data is desired. 

Test-time adaptation~\cite{tta_1,tta_2,tta_3,tta_4} emerges as a promising solution that tackles the issue of distribution shifts by fine-tuning the models during the testing phase. Different from the aforementioned methods, test-time adaptation does not require exploring additional atomic structures or extra training data, instead opting for on-the-fly model adjustments to the model in response to the characteristic of test data with only a modest increase in computational overhead. Test-time adaptation has been proven effective in various domains such as image classification~\cite{tta_3,tta_4,para_free_tta,Continual_tta}, semantic segmentation~\cite{tta_seg,tta_seg2,tta_seg3,tta_seg4}, and object detection~\cite{tta_det,tta_det2,tta_det3,tta_det4}. Nevertheless, its application in predicting interatomic potentials remains unexplored. For a successful implementation of test-time adaptation in MLIP, it is crucial to devise task-specific strategies which account for the specific characteristics of atomic structure data. A well-crafted test-time adaptation tailored for MLIP holds substantial promise in improving the accuracy of MLIP, as well as the stability and reliability of MD simulations driven by MLIPs.

In this work, we propose an online Test-time Adaptation strategy for Interatomic Potentials (TAIP) aimed at mitigating the impact of distribution shifts on MLIP applications. We design a dual-level self-supervised learning scheme that help extract both global and local structural information using an encoder. During training, the encoder is trained by the combined losses from both MLIP and self-supervised learning tasks. At the inference stage, the encoder is updated once per test sample by minimizing the self-supervised learning loss, subsequently yielding the final energy and force predictions. This fine-tuning process during inference allows the encoder to extract more adaptive features for test data. We test the accuracy of MLIPs on four datasets, including MD17, ISO17, water, and electrolyte solutions. Compared to baselines, TAIP reduces the prediction errors by an average of 30\% without using any additional data. Moreover, we assess the influence of TAIP on the MD simulation stability using periodic water and electrolyte solution datasets and find that TAIP enables stable MD simulations throughout even under conditions where baselines collapse. Finally, visual analysis of feature distributions confirms that TAIP curtails the distribution shifts between training and test datasets.

\section{Results}\label{sec2}

\subsection{Overview of TAIP} 

\begin{figure}[tbp]
    \begin{center}
  \subfigure{\label{fig:1_a}}
  \subfigure{\label{fig:1_b}}
    \begin{overpic}[width=0.95\textwidth]{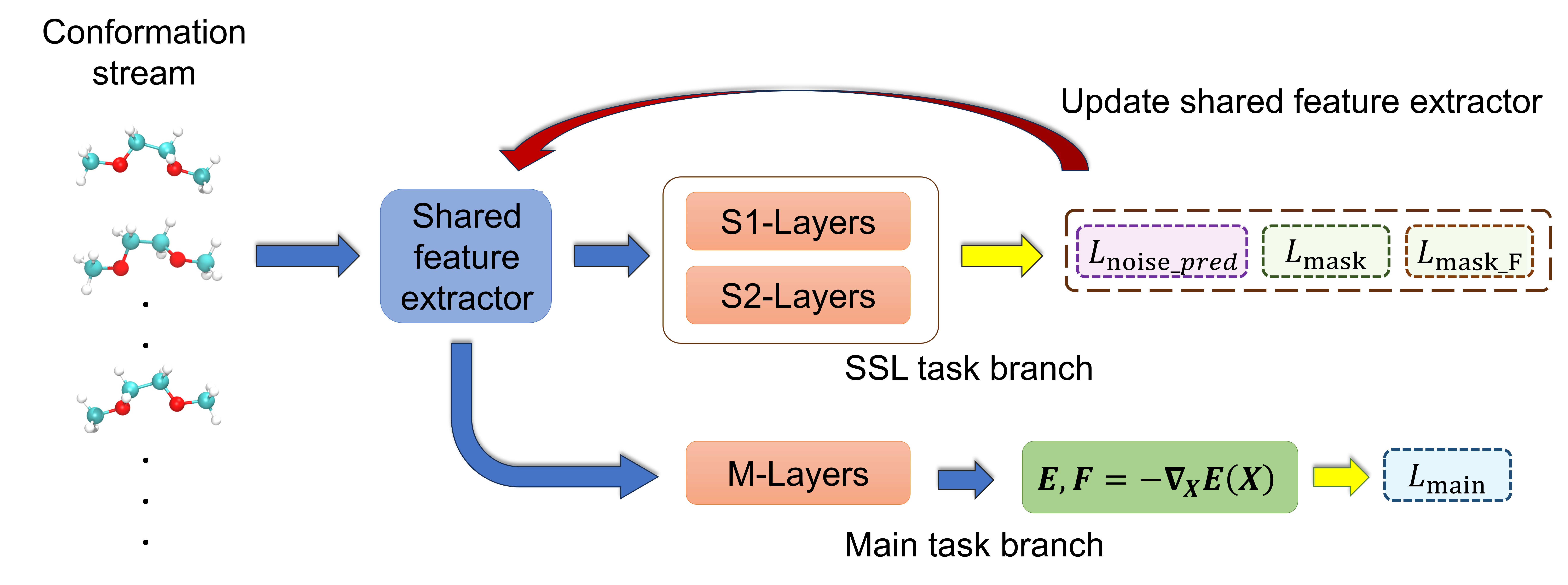}
			\put(-7,20){\textbf{(a)}}
	\end{overpic}
    \begin{overpic}[width=0.95\textwidth]{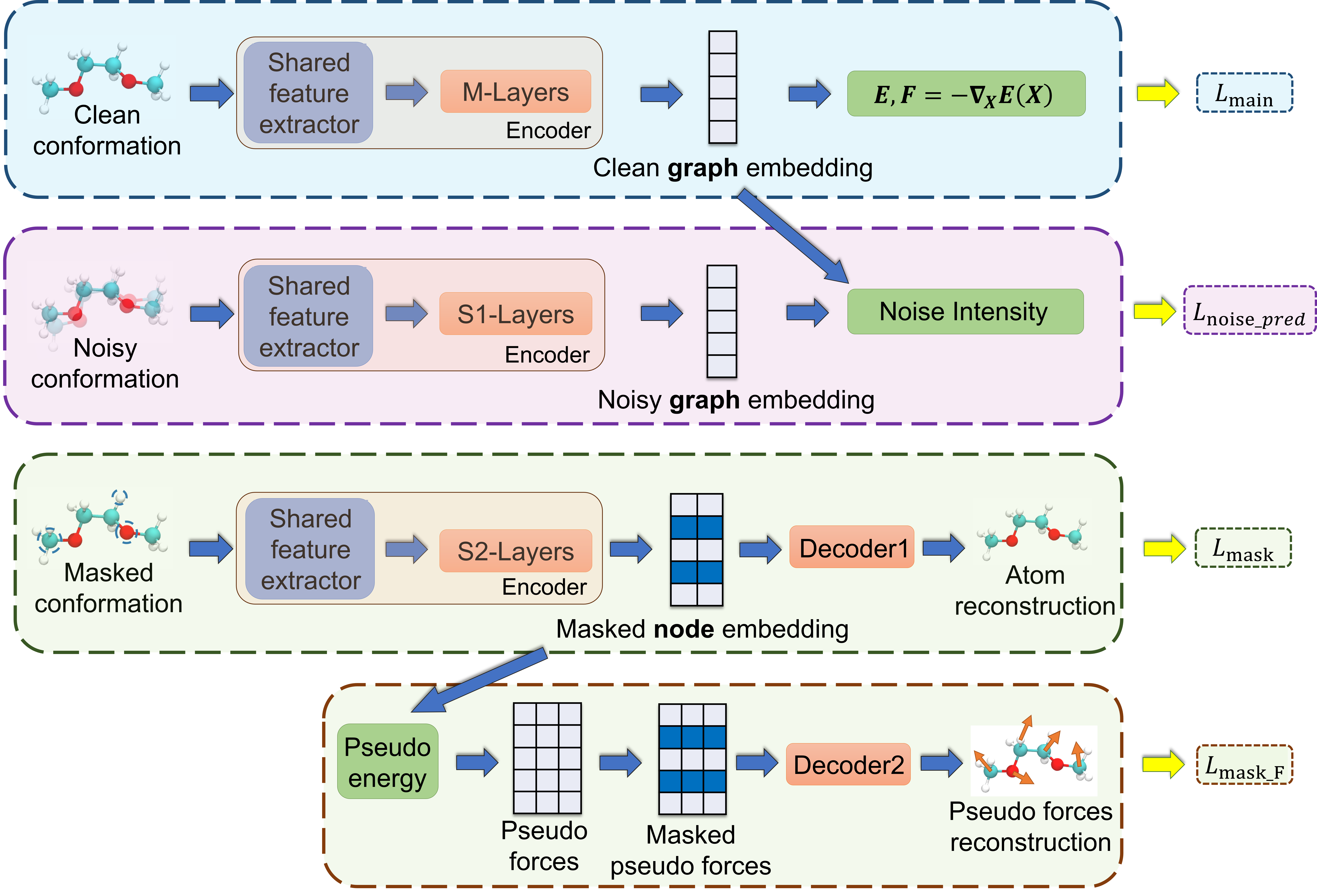}
			\put(-7,45){\textbf{(b)}}
	\end{overpic}
    \end{center}
	\caption{\textbf{Overview.} (a) TAIP inlcudes three stages: training, test-time adaptation, and inference. The supervised and self-supervised learning losses are minimized simultaneously in the training phase. The encoder is updated once per test sample in the adaptation phase. (b) Details of the tasks. The Main task (block 1) aims to predict energy and force. The self-supervised learning tasks (block 2-4) involve noise intensity prediction, atom feature reconstruction, and pseudo force reconstruction.}
	\label{fig:pipeline}
\end{figure}

Three stages of TAIP include training, test-time adaptation, and inference, as shown in \cref{fig:1_a}. In the training phase, the network parameters are updated according to the supervised learning loss of the main task, namely energy and force prediction, as well as three self-supervised learning losses. In the test-time adaptation phase, we first input the test sample into the encoder and update its parameters once per test sample to reduce the self-supervised learning losses. This unique process facilitates the model's adaptation to the test data, thereby raising its generalization capabilities on test samples. Since the parameter update during adaptation is performed only once, the computational time and cost will not increase substantially. At the inference stage, the test sample passes through the encoder to predict the energy and forces, which is updated during the adaptation phase. 

The \textbf{main task} shown in block 1 of \cref{fig:1_b} is a typical MLIP task that predicts energy and force. We use a graph to represent a molecular conformation and feed it into the encoder to generate the graph feature. The graph feature is then utilized to predict the potential energy of the molecule, and the force exerted on each atom is calculated by taking the negative gradient of the potential energy.

The \textbf{noise intensity prediction task} is shown in block 2 of \cref{fig:1_b}. For each sample, we perturb the coordinates of atoms with random noises drawn from a Gaussian distribution, and the variance is defined as the noise intensity. The encoder takes the perturbed structure as input and produces the noisy graph feature. Subsequently, the noisy graph feature is combined with the original clean graph feature to jointly estimate the noise intensity, which reflects the dissimilarity between noisy and clean data. Instead of directly predicting the precise value of noise, we categorize the intensity into several bins and determine the bin to which the feature corresponds. This approach naturally groups similar configurations while separating dissimilar ones and thus reduces abrupt changes or outlines, leading to a smoother configuration representation space. By focusing on the key features that distinguish between different levels of noise, the model can learn more robust features and enhance its generalization ability to new scenarios. 

The \textbf{atom feature recovery task} is shown in block 3 of \cref{fig:1_b}. We randomly mask some atoms in each sample with a certain ratio and then obtain the atomic representations by feeding the masked atomic structure to the encoder. The representations are subsequently used to rebuild the original atomic features. %We then use them to rebuild the atomic features. 
We can infer the most probable types of atoms in the vicinity of each atom by restoring the masked atoms, and thereby improving the extraction of local information.

The \textbf{pseudo force recovery task} is shown in block 4 of \cref{fig:1_b}. We mask the pseudo forces, which are defined as the negative gradient of the pseudo energy obtained from the masked node features. Then, we feed the masked pseudo forces into the decoder to reconstruct pseudo forces. By restoring the pseudo forces, the model can capture the relationship among the forces on neighboring atoms, which can better understand the local environment of the atoms and is helpful for the prediction of the atomic forces.

%In the training phase, the parameters of the whole network are updated according to the supervised loss of the main task, namely energy and force predictions, and the losses from dual-level SSL tasks, including predicting noise intensity and reconstructing masked atoms and pseudo forces. In the inference phase, we first input the test sample into the SSL task branch and update the parameters of the shared feature extractor once per test sample to reduce the SSL loss. Afterward, the test sample passes through the main task branch to predict the energy and forces. This unique process narrows the gap between the training and test data, thereby raising the model's generalization capabilities on test samples. Since the fine-tuning operation during inference is performed only once, the computational time and cost for inference will not increase substantially.

\subsection{Experiments}\label{sec3}

We conducted a series of experiments across multiple datasets to evaluate the performance of TAIP on the widely used MLIPs of SchNet~\cite{schnet} and PaiNN~\cite{painn}, which are invariant and equivariant, respectively. First, we tested the enhanced accuracy in predicting energies and forces on the MD-17 dataset~\cite{chmiela2017machine}. Next, we extended our evaluation to the ISO17~\cite{schnet} dataset and complex systems featuring periodic boundary conditions, including water and electrolyte solutions. The test data from these datasets were designed to have a larger distribution shift from the training data, as indicated in Supplementary Figure S1, and therefore are used to investigate the improvement in generalizability achieved by TAIP. 
We then performed MD simulations with MLIPs on the aforementioned complex systems to assess the practical utility of TAIP.
Moreover, we explored the impact of TAIP on feature distributions through dimensionality reductions using the t-SNE method~\cite{tsne}, illustrating the mechanism behind the improvement from the feature embedding perspective.

\textbf{MD-17.} We evaluated the performance of TAIP on the widely used MD-17 dataset, which contains small organic molecules with reference values of energies and forces generated by ab initio MD simulations. Table~\ref{tab:md17} compares the mean absolute errors (MAEs) of the predicted energies and forces for different molecules using different models. The results show that TAIP-based models remarkably outperform their corresponding baseline models for all eight molecules, demonstrating the generalization ability of our method. 

\begin{table*}
  \caption{
    \textbf{Results on MD17 dataset.} Energy and force MAEs are reported in units of kcal/mol and kcal/mol/\AA, respectively. Standard deviations are calculated from five independent experiments. Results of all baseline models are directly taken or adapted (if the unit varies) from the original papers and standard deviations are not provided. The MAEs of TAIP-based models are consistently less than the corresponding baselines and are shown in bold.}
  \centering
  \label{tab:md17}
  \resizebox{\textwidth}{!}{
    \begin{tabular}{@{}lllllllllll  @{}}
      \toprule
       &    & SchNet & SchNet-TAIP                       &  &   & PaiNN  & PaiNN-TAIP  &    &   \\            
      \midrule
      \multirow{2}{*}{\textbf{Aspirin}}
       & Energy                                & 0.37                        & \textbf{0.322}$\pm$\textbf{0.041}                                       &              &    & 0.159 & \textbf{0.144}$\pm$\textbf{0.011}&&                                        \\
       & Force                                & 1.35                        & \textbf{0.853}$\pm$\textbf{0.045}                                       &              &    & 0.376& \textbf{0.330}$\pm$\textbf{0.012}&& \\
    \midrule
      \multirow{2}{*}{\textbf{Benzene}}
       & Energy                                & 0.08                       & \textbf{0.059}$\pm$\textbf{0.019}                                      &          &       & 0.101&\textbf{0.067}$\pm$\textbf{0.012}& &                 \\
       & Force                                & 0.31                        & \textbf{0.251}$\pm$\textbf{0.013}                                     &          &       & 0.226&\textbf{0.186}$\pm$\textbf{0.007}& &                 \\       
    \midrule

        \multirow{2}{*}{\textbf{Ethanol}}
       & Energy                              & 0.08                       & \textbf{0.057}$\pm$\textbf{0.020}                                      &             &    & 0.086&\textbf{0.061}$\pm$\textbf{0.015}&&                                        \\
& Force                                & 0.39                        & \textbf{0.272}$\pm$\textbf{0.035}                                      &           &    & 0.230&\textbf{0.180}$\pm$\textbf{0.010}&&  \\
    \midrule
      \multirow{2}{*}{\textbf{Malonaldehyde}}
       & Energy                               & 0.13                        & \textbf{0.118}$\pm$\textbf{0.011}                                     &           &       & 0.100&\textbf{0.091}$\pm$\textbf{0.011}&&
       \\
       & Force                                 & 0.67                        & \textbf{0.612}$\pm$\textbf{0.023}                                     &          &       & 0.319& \textbf{0.297}$\pm$\textbf{0.009}&&
       \\

    \midrule
      \multirow{2}{*}{\textbf{Naphthalene}}
       & Energy                                & 0.16                        & \textbf{0.110}$\pm$\textbf{0.017}                                                   &    & &0.113&\textbf{0.093}$\pm$\textbf{0.005}&&                                       \\
       & Force                               & 0.58                        & \textbf{0.318}$\pm$\textbf{0.033}                                   &             &    & 0.079&\textbf{0.072}$\pm$\textbf{0.003}&&                                       \\
    \midrule
      \multirow{2}{*}{\textbf{Salicylic acid}}
       & Energy                                & 0.20                        & \textbf{0.154}$\pm$\textbf{0.026}                                     &            &    & 0.114&\textbf{0.105}$\pm$\textbf{0.004}&&                                       \\
       & Force                                & 0.85                        & \textbf{0.671}$\pm$\textbf{0.036}                                     &             &    & 0.209&\textbf{0.193}$\pm$\textbf{0.003}&&                                       \\
    \midrule
      \multirow{2}{*}{\textbf{Tolunene}}
       & Energy                               & 0.12                        & \textbf{0.103}$\pm$\textbf{0.010}                                      &              &    & 0.119&\textbf{0.109}$\pm$\textbf{0.003}&&                                       \\
       & Force                                & 0.57                        & \textbf{0.461}$\pm$\textbf{0.023}                                     &             &    & 0.102&\textbf{0.090}$\pm$\textbf{0.004}&& \\
    \midrule
      \multirow{2}{*}{\textbf{Uracil}}
       & Energy                               & 0.14                        & \textbf{0.109}$\pm$\textbf{0.006}                                                   & &   & 0.104&\textbf{0.090}$\pm$\textbf{0.006}&&                                        \\
       & Force                                & 0.56                        & \textbf{0.493}$\pm$\textbf{0.013}                                      &              &    & 0.143&\textbf{0.130}$\pm$\textbf{0.003}&&   \\
    % \midrule
    %   \multirow{1}{*}{\textbf{std. MAE}}
    %    & F & 1.11                                & 2.38                        & \textbf{1.08}                                       & 1.10              & 0.79   & 0.97&0.85&&                                        \\
      \bottomrule
    \end{tabular}
  }
\end{table*}

\textbf{ISO17.} 
The ISO17 dataset contains trajectories of isomers of C$_7$O$_2$H$_{10}$. We consider two scenarios with the ISO17 dataset. In the first scenario (known molecules/unknown conformation), the isomers in the test set are also present in the training set. In the second scenario (unknown molecules/unknown conformation), the test set contains a different subset of isomers. This task is more challenging and is used to evaluate TAIP in the case where training and test data are drawn from different distributions. Table \ref{tab:isomer} shows that, for the energy and force MAEs on the ISO17 dataset, TAIP achieves an average reduction of 40\% and 31\% in scenario 1 and scenario 2, respectively.

\begin{table*}
\caption{\textbf{Results on ISO17, water, and electrolyte datasets.} These datasets are used to evaluate TAIP in the case of large distribution shift. For ISO17 dataset, energy and force MAEs are reported in units of kcal/mol and kcal/mol/\AA, respectively. For water and electrolyte dataset, energy and force MAEs are reported in units of eV and eV/\AA, respectively. Standard deviations are calculated from five independent experiments. The standard deviations for ISO17 dataset with baseline models are not provided in the original papers. The MAEs of TAIP-based models are consistently less than the corresponding baselines and are shown in bold.}\label{tab:isomer}
\centering
\small

\resizebox{\textwidth}{!}{
\begin{tabular}{@{}lllllllllll  @{}}
\toprule
& &  & \text{SchNet}&\text{SchNet-TAIP}&\text{PaiNN}&\text{PaiNN-TAIP} \\

 \midrule
\multirow{2}{*}{\text{ISO17}}
& \text{known molecules/} &   \text{Energy}   &  0.39 &\textbf{0.291}$\pm$\textbf{0.013} &0.32&\textbf{0.111}$\pm$\textbf{0.079}\\
& \text{unknown conformation} & \text{Force} &    1.00 &\textbf{0.554}$\pm$\textbf{0.079} &0.26&\textbf{0.202}$\pm$\textbf{0.011}\\ \midrule
\multirow{2}{*}{\text{ISO17}}
& \text{unknown molecules/} & \text{Energy}   &    2.40 &\textbf{1.130}$\pm$\textbf{0.123} &0.92& \textbf{0.662}$\pm$\textbf{0.054}\\
& \text{unknown conformation} & \text{Force} &    2.18 &\textbf{1.783}$\pm$\textbf{0.111} &0.89&\textbf{0.684}$\pm$\textbf{0.048}\\

 \midrule
\multirow{2}{*}{\text{Water}}
& \text{known phase/} &   \text{Energy} &    0.283$\pm$0.027  & \textbf{0.236}$\pm$\textbf{0.023} &0.216$\pm$0.034&\textbf{0.090}$\pm$\textbf{0.017}&\\
& \text{unknown configuration} & \text{Force} & 0.076$\pm$0.005 & \textbf{0.066}$\pm$\textbf{0.004}  &0.027$\pm$0.004 &\textbf{0.021}$\pm$\textbf{0.004}&\\ \midrule
\multirow{2}{*}{\text{Water}}
& \text{unknown phase/}& \text{Energy}    & 1.193$\pm$0.237 & \textbf{0.423}$\pm$\textbf{0.204} &0.585$\pm$0.233&\textbf{0.042}$\pm$\textbf{0.084}&\\
& \text{unknown configuration} & \text{Force} &   0.083$\pm$0.010 & \textbf{0.055}$\pm$\textbf{0.005}&0.023$\pm$0.003&\textbf{0.018}$\pm$\textbf{0.003}&\\
\bottomrule

\multirow{2}{*}{\text{Electrolyte}}
& \text{known concentration/} &   \text{Energy}   &    0.263$\pm$0.031 &\textbf{0.203}$\pm$\textbf{0.010} &0.238$\pm$0.014&\textbf{0.199}$\pm$\textbf{0.006}\\
& \text{unknown configuration} & \text{Force} &    0.064$\pm$0.012 &\textbf{0.043}$\pm$\textbf{0.008} &0.049$\pm$0.010&\textbf{0.037}$\pm$\textbf{0.003}\\ \midrule
\multirow{2}{*}{\text{Electrolyte}}
& \text{unknown concentration/} & \text{Energy}   &    9.459$\pm$0.459&\textbf{4.816}$\pm$\textbf{0.274} &9.872$\pm$0.423& \textbf{4.914}$\pm$\textbf{0.307}\\
& \text{unknown configuration} & \text{Force} &    0.341$\pm$0.043 &\textbf{0.197}$\pm$\textbf{0.032} &0.143$\pm$0.025&\textbf{0.113}$\pm$\textbf{0.012}\\
\bottomrule
\end{tabular}
}
\end{table*}

\begin{figure}[t]
    \centering
    % \subfigure{\label{fig:2_a}}
    % \subfigure{\label{fig:2_b}}
    % \subfigure{\label{fig:2_c}}
    % \subfigure{\label{fig:2_d}}
    \subfigure[]{\includegraphics[height=0.24\textwidth]{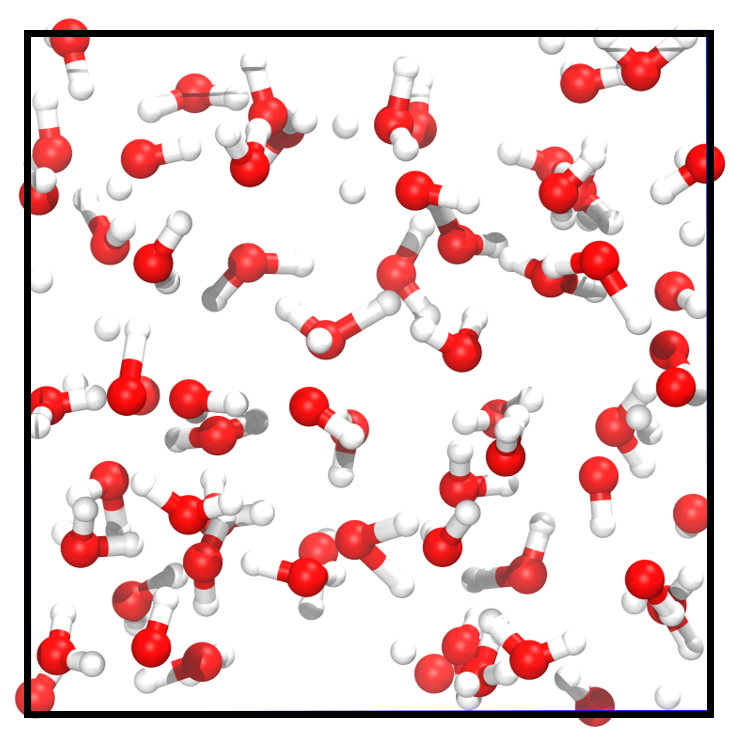} \label{fig:2_a}}
    \subfigure[]{\includegraphics[height=0.24\textwidth]{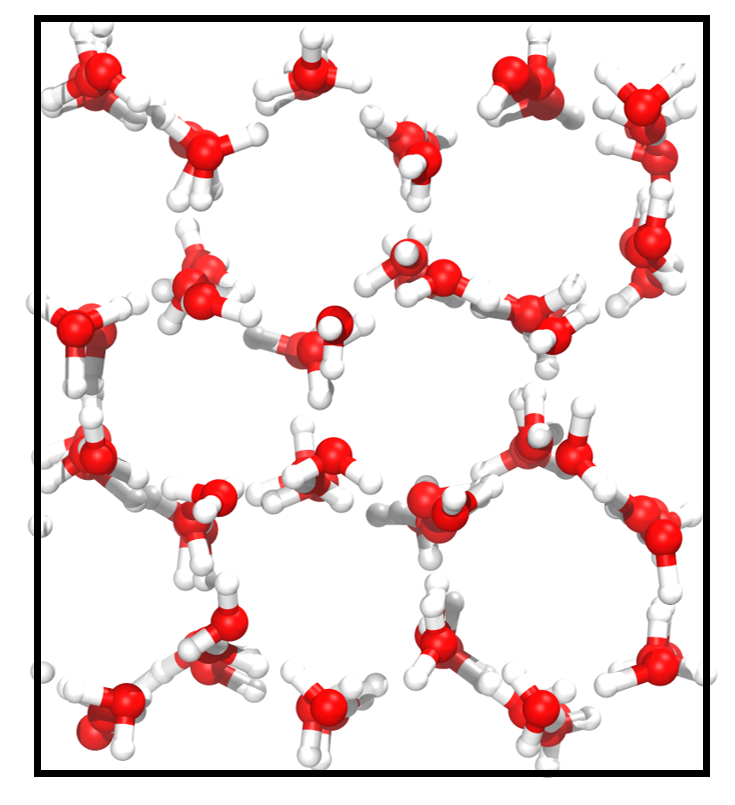} \label{fig:2_b}}
    \subfigure[]{\includegraphics[height=0.24\textwidth]{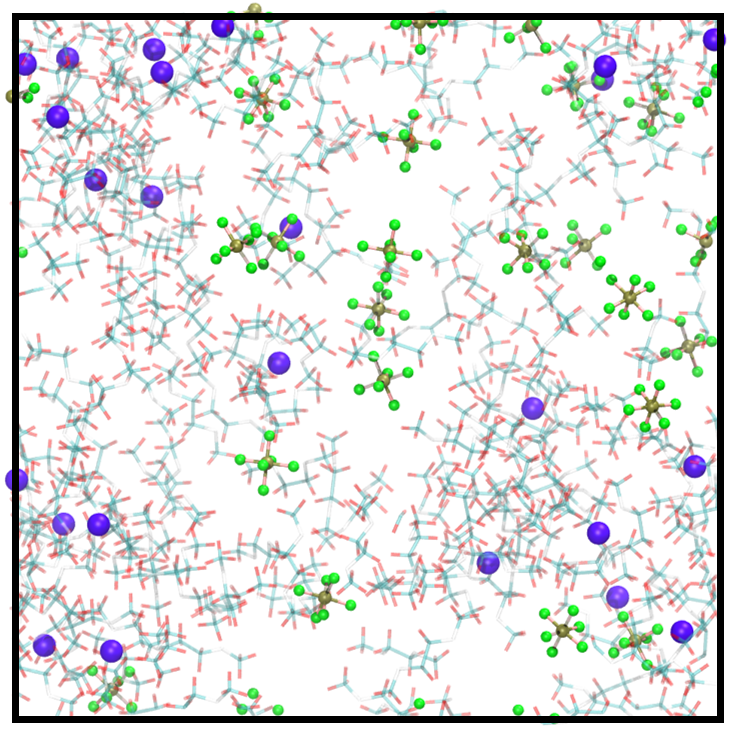} \label{fig:2_c}}
    \subfigure[]{\includegraphics[height=0.24\textwidth]{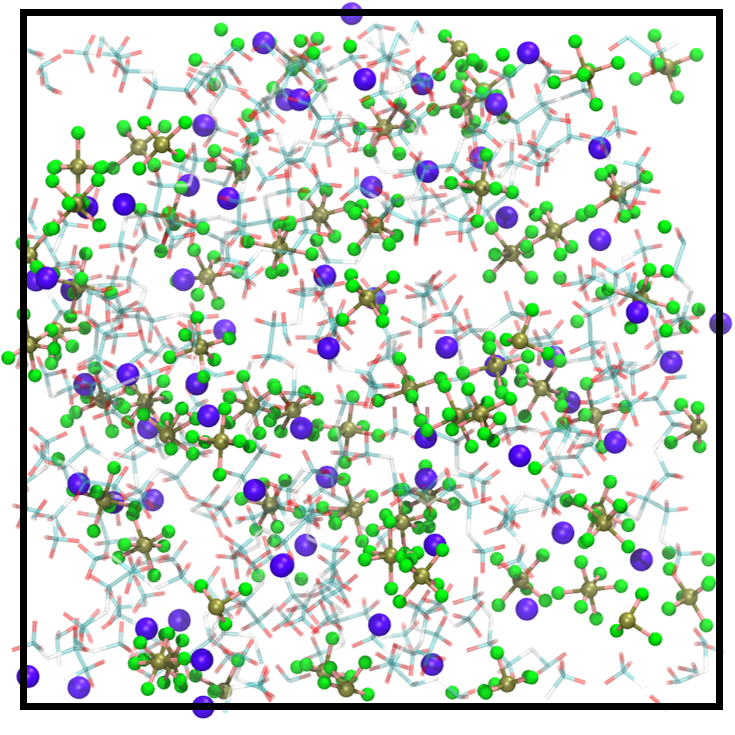} \label{fig:2_d}}
    
    \caption{\textbf{Visualization of atomic structures.} (a) Liquid Water. (b) Ice. (c) 1 M electrolyte solutions. (d) 4 M electrolyte solutions. Color scheme: H-white; O-red; Li-purple; P-ochre; F-green. DME solvents are displayed in line style to highlight the difference between 1 M and 4 M solutions.}
    \label{fig:dataset}
    \vspace{-3.0ex}
\end{figure}

\textbf{Liquid Water and Ice.} 
Water has complex phase behaviors that pose considerable challenges for computational studies. As shown in \cref{fig:2_a} and \cref{fig:2_b}, there exists a substantial difference between the microscopic structures of liquid water and ice. In the liquid state, water molecules form a highly dynamic network through hydrogen bonding. In contrast, water molecules form a stable hexagonal lattice structure in an ice crystal. The structural differences between different phases can lead to a decrease in prediction accuracy. Thus, we train the models only with liquid water, using a training set of 1000 frames and a validation set of 100 frames, and report the test accuracy on randomly sampled 1,000 liquid water and ice structures from the remaining dataset, respectively. The results of MAE on liquid water (known phase/unknown configuration) and on ice (unknown phase/unknown configuration) are shown in \cref{tab:isomer}. The TAIP method notably reduces the errors of force and energy predictions by an average of 40\% for unknown configurations across known and unknown phases in the two baseline models.

\textbf{Electrolyte solutions of different concentrations.} We examine TAIP on the electrolyte solution dataset, which is developed in our previous work~\cite{cui2024geometry}, to further analyze compatibility with complex systems. Consisting of charged ions and electrolyte solvents, the dataset includes more elements, stronger electrostatic forces, and more complicated interatomic interactions, thereby exhibiting a higher degree of complexity than water. Training MLIPs for such complex systems and generalizing the trained models on different concentrations are both challenging.
Here, we train the models on the 1 M electrolyte solutions and test the generalization across configurational space at different concentrations. The atomic structures of 1 M and 4 M electrolyte solutions are illustrated in \cref{fig:2_c} and \cref{fig:2_d}, respectively. 
The MAE results for the test set of 1 M solutions (known concentration/unknown configuration) and 4 M solutions (unknown concentration/unknown configuration) are presented in \cref{tab:isomer}. The data reveals that TAIP substantially enhances the precision of energy and force predictions for unknown configurations, irrespective of whether the concentration is known or unknown. 
 
\textbf{MD simulations.}
The MAEs of predicted forces and potential energies on fixed datasets are not enough to characterize the performance of MLIPs on long-time MD simulations~\cite{fu2022forces}. During MD simulations, new molecular structures are constantly produced, and the false prediction of the interatomic forces on a single atom may cause the collapse of the entire simulation system. Therefore, we perform MD simulations using SchNet with and without TAIP on periodic systems (water and electrolyte solutions) to further analyze how TAIP can affect the predicted forces and, thus, the quality of MD simulations.

\begin{figure}[t]
    \begin{center}
    \begin{overpic}[width=0.9\textwidth]{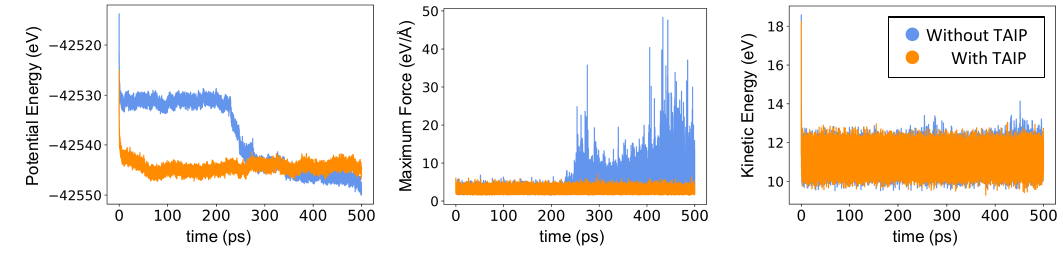}
			\put(-5,15){\textbf{(a)}}
	\end{overpic}
    \begin{overpic}[width=0.9\textwidth]{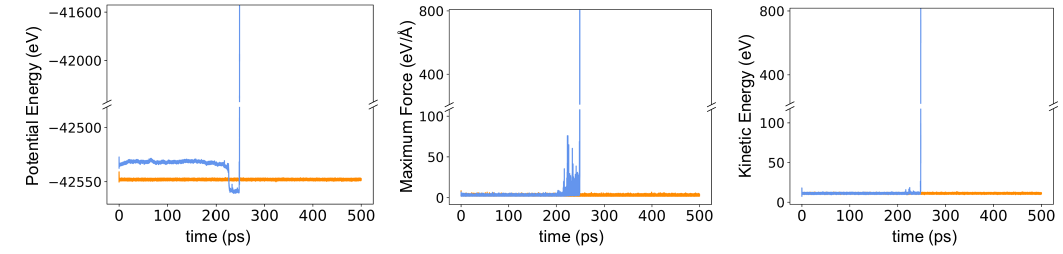}
			\put(-5,15){\textbf{(b)}}
	\end{overpic}
    \begin{overpic}[width=0.9\textwidth]{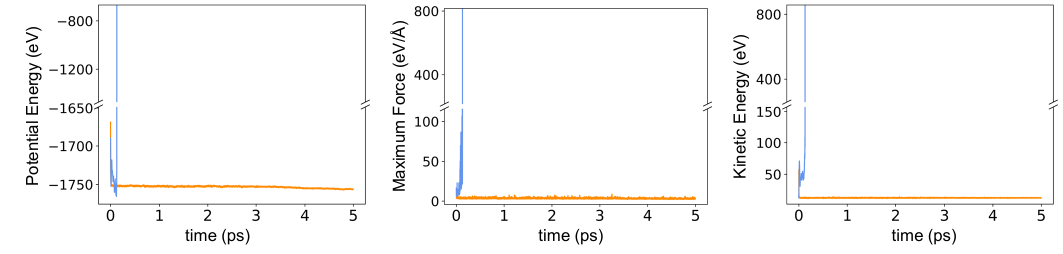}
			\put(-5,15){\textbf{(c)}}
	\end{overpic}
    \begin{overpic}[width=0.9\textwidth]{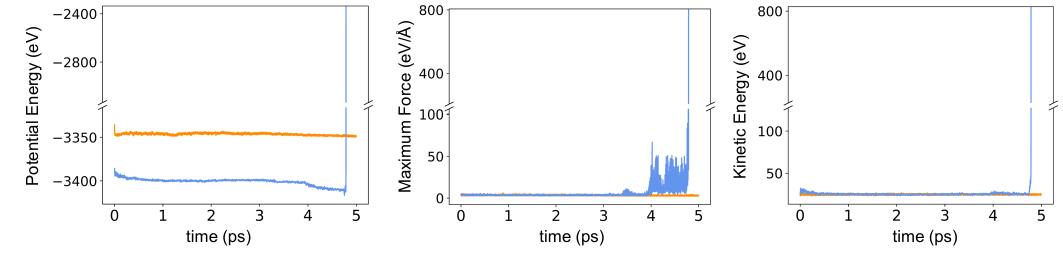}
			\put(-5,15){\textbf{(d)}}
	\end{overpic}
    \end{center}
    \caption{\textbf{Results of MD simulations.} (a) Liquid water. (b) ice. (c) 1M electrolyte solutions. (d) 4M electrolyte solutions. The MD simulations with baseline MLIPs (blue) breakdown in a short time, whereas those with TAIP (yellow) are stable throughout the simulation time.}
    \label{fig:MD}
    \vspace{-3.0ex}
\end{figure}

All simulations are set up with a timestep of 0.5 fs, using the Berendsen thermostat~\cite{berendsen1984molecular} as the temperature coupling method with a coupling temperature of 300 K and a decaying time constant ${\tau}$ of 100 fs. Without TAIP, the simulations for electrolyte solutions with concentrations of 1 M and 4 M collapse due to losing several atoms at the step of 258 ($\sim$0.1 ps) and of 1843 ($\sim$0.9 ps), respectively. The simulation for water exhibits undesired force and energy instability after 250 ps, and that for ice collapses at 248 ps. In contrast, all the simulations with TAIP can run stably for more than 500 ps. 
The simulation trajectories are provided in Supplementary Figure S2 and Supplementary Video S1-S4. 
We show the potential energies, kinetic energies, and the maximum absolute force of these trajectories on \cref{fig:MD}. It is evident that the failure of MD simulations is caused by the divergence of predicted potential energies and forces. Instability is a fundamental issue for MLIP-based MD simulations, as the accuracy of MLIPs without TAIP is crucially compromised when encountered with unknown molecular configurations. The inclusion of the TAIP method remarkably enhances the overall performance of MILP-based MD simulations when dealing with unknown molecular structures, and it can be suggested as an applicable solution to MD instability. 

% We also compare the radial distribution functions (RDFs) between key atoms in MD trajectories using MLIPs with TAIP against AIMD trajectories on \cref{fig:rdf}. It can be indicated that the MLIP-MD simulations can replicate most of the structural information generated by AIMD high-accuracy simulations. This further signifies this approach of increasing the accuracy and generalization of MLIPs. 

% \begin{figure}[t]
%     \centering
%     \includegraphics[width=0.50\linewidth]{tta_rdf.png}
%     \caption{\textbf{Radial Distribution Functions} RDFs (a) between Na and O in electrolyte solutions and (b) between O and O in liquid water.}
%     \label{fig:rdf}
%     \vspace{-3.0ex}
% \end{figure}

\begin{figure}[t]
    \centering
    \includegraphics[width=0.8\linewidth]{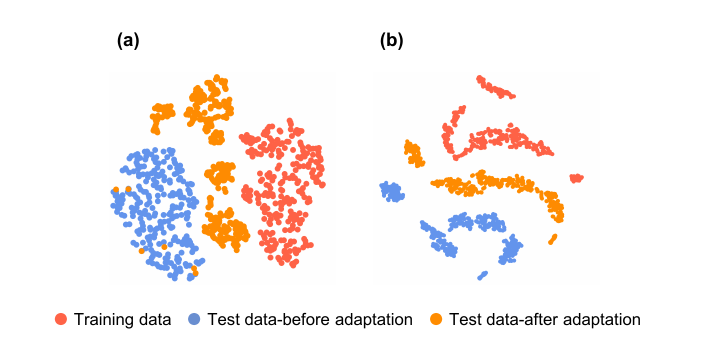}
    \caption{\textbf{Visualization of feature space.} (a) Liquid water (train) and ice (test). (b) 1 M (train) and 4 M (test) electrolyte solutions. The feature embeddings of test data are closer to those of training data.}
    \label{fig:visual}
    \vspace{-3.0ex}
\end{figure}

\textbf{Feature visualization.}
To visually identify the effect of TAIP. We perform a series of dimensionality reductions using the t-SNE method~\cite{tsne} to map the feature embeddings of the training dataset and the testing dataset on benchmark models and TAIP models. In \cref{fig:visual}, we can notice that the feature embeddings of the test dataset in the TAIP models are much closer to that of the training dataset than the feature embeddings in the baseline models. It indicates that the TAIP enables the MLIPs to process unknown molecular structures with less uncertainty, which helps the energies and forces predictions to harness more accuracy and generalizability. 

\section{Discussion}\label{sec3}

The primary challenge for MLIPs in practical applications lies in their ability to make accurate predictions on test data that exhibit a distribution shift from the training set. Specifically, a feasible MLIP must maintain sufficient accuracy for the new atomic structures that continuously emerge during simulations. This study demonstrates that our TAIP framework remarkably enhances the performance of MLIPs on test datasets, which are designed to exhibit substantial distribution shifts from the training data, without requiring any extra data. The efficacy of TAIP is attributed to self-supervised learning tasks including noise intensity prediction, atomic feature recovery, and pseudo force recovery. These tasks are trained concurrently with the interatomic potential task during the training phase. During the inference phase, the shared feature extractor is refined through self-supervised learning tasks in an online manner. This approach effectively bridges the domain gap between training and testing datasets, as evidenced by t-SNE visualizations of the feature space. Beyond improving the accuracy of predicting energy and force, TAIP also enables stable MLIP-based MD simulations, outperforming baseline models that otherwise fail. The experiments across various molecular systems underscores the vast potential of our method.

\backmatter

\section{Methods}\label{sec4}

\subsection{Dataset}

\textbf{MD-17.} 
The MD17 dataset is composed of ab initio molecular dynamics trajectories of eight small molecules \cite{chmiela2017machine}. Obtained from \url{http://www.sgdml.org/#datasets}, we use the training set of 1,000, the validation set of 1,000, and the test set of 1,000 configurations for each small molecule.

\textbf{ISO17.} 
ISO17 dataset consists of ab initio MD trajectories of 129 isomers whose energies and forces are calculated by Density Functional Theory (DFT)~\cite{schnet}. We adopted the same splitting strategy reported in the original literature\cite{schnet}, using 80\% trajectory steps of the 80\% isomers for training and validation (totaling 404,000 molecular conformations for training and 4,000 for validation). We further set up two test datasets to evaluate the effectiveness of TAIP: (1) The other 20\% unseen trajectory steps of the isomers included in the training set (totaling 101,000 conformations) and (2) all of the remaining 20\% isomers not included in the training and validation set (totaling 130, 000 conformations).

\textbf{Liquid Water and ice.}
The atomic structures are sampled from MD trajectories, which consist of a training set of 1,000 snapshots, a validation set of 500 snapshots, and a test set of 1,000 liquid water. In addition, 1,000 ice snapshots in the form of hexagonal ice crystals are sampled to be the other test set. There are 96 molecules or 288 atoms in each snapshot.
The forces and energies for each sampled snapshot are calculated by DFT using the cp2k package\cite{kuhne2020cp2k}. The DFT calculations are conducted using PBE exchange-correlation functional\cite{perdew1996generalized} with the Projector Augmented-Wave (PAW) pseudo-potential\cite{blochl1994projector}. The DFT-D3 method is used for the vdW-dispersion energy correction\cite{grimme2010consistent}. A single gamma k-point is used to sample the Brillouin zone, the cutoff energy for the plane-wave-basis set is set to be 400 eV, and the electronic self-consistency is considered to be achieved if the change of total energy between two steps is smaller than $10^{-6}$ eV. 

\textbf{Electrolyte solutions.} 
The electrolyte solutions dataset is taken from our previous work~\cite{cui2024geometry}. Here, we used the combination of [$\rm Li^{+}$$\rm PF_{6}^{-}$DME], $[\rm Na^{+}$$\rm PF_{6}^{-}$DME], $[\rm Li^{+}$$\rm Tf_{2}N^{-}$DME], $[\rm Na^{+}$$\rm Tf_{2}N^{-}$DME], [$\rm Li^{+}$$\rm PF_{6}^{-}$EC+DMC], $[\rm Na^{+}$$\rm PF_{6}^{-}$EC+DMC], $[\rm Li^{+}$$\rm Tf_{2}N^{-}$EC+DMC], and $[\rm Na^{+}$$\rm Tf_{2}N^{-}$EC+DMC] as electrolyte solutions with ionic concentrations of 1 M and 4 M. 
The training and validation data are randomly selected from the 1 M solutions. The training set contains 1000 samples and the validation set contains 250 samples. We construct two different test sets to evaluate TAIP: one is randomly sampled from the remaining 1 M solutions and the other is sampled from the 4 M solutions. 

\subsection{Training Settings}

The TAIP framework is implemented based on PyTorch 1.8.0. The experiments are conducted with NVIDIA GeForce RTX 3090 GPU. The models are trained using the Adam optimizer~\cite{kingma2014adam}, employing single-GPU training for efficient processing. The hyperparameters are provided in Supplementary Table S1-S3. 

In order to achieve optimal performance on both properties, we incorporate both energy $E$ and forces $\mathbf{F}_i$ into the criterion for training:

\begin{equation}
    \mathcal{L}_{\text{all}} = \left|{E}-{\hat{E}}\right|^{2}+\frac{\lambda}{n}\sum_{i=1}^{n}\left\|\mathbf{F}_{i}-\left(-\frac{\partial{\hat{E}}}{\partial\mathbf{R}_{i}}\right)\right\|^{2} + \mathcal{L}_{\text{ni}} + \mathcal{L}_{\text{ma}}+ \mathcal{L}_{\text{mf}},
\end{equation}

\noindent where $\hat{E}$ represents the predicted energy, $n$ is the number of atoms in each sample and $\mathbf{R}_{i}$ represents the coordinate of atom $i$. The weight $\lambda=100$ is used to in line with the setting of the baseline models. The $\mathcal{L}_{\text{ni}}$, $\mathcal{L}_{\text{ma}}$ and $\mathcal{L}_{\text{mf}}$ are self-supervised learning losses corresponding to noise intensity prediction, masked atom reconstruction and masked pseudo force reconstruction, respectively. 

For the periodic systems including water and electrolyte solutions, we employ the atomization energy as the target energy by shifting the original potential energy according to the energy of each single atom:

\begin{equation}
    E^\text{shift} = E - \sum_{i=1}^n{E_i^\text{ref}}
\end{equation}

\noindent where ${E_i^\text{ref}}$ is the potential energy of a single atom $i$ in vacuum.

\textbf{Noise intensity prediction.}
For each sample, we randomly choose a noise intensity $l$ from a uniform distribution $U(1,L)$ and perturb the coordinates accordingly, which will facilitate the exploration of the configurational space. The GNNs are trained to predict the noise intensity instead of the specific noise value. The graph feature of the perturbed structure $\mathbf{\hat{u}}$ is concatenated with that of the original structure $\mathbf{u}$ and passes through a classification head $\mathbf{\Phi}_{\text{scale}}$ to produce the classification logit. The training objective can be formulated as follows:

\begin{equation}
\mathcal{L}_{\text{ni}} = 
\mathcal{L}_{\text{CE}}\left(\mathbf{I}(l),\mathbf{\Phi}_{\text{scale}}(\mathbf{u}||\mathbf{\hat{u}})\right),
\end{equation}

\noindent where $\mathbf{I}(l)$ is the one-hot encoding of $l$, and $\mathcal{L}_{\text{CE}}$ denotes the cross entropy loss.

\textbf{Atomic feature recovery.}
The masked atomic structures are fed into the encoder to get the masked node features. We utilize the cosine error as the criterion for reconstructing the original node features. Additionally, we reduce the weight of easy samples in the training set by scaling the cosine error with a power of $\gamma$, as shown below: 

\begin{equation}
\mathcal{L}_{\text{ma}}=\frac{1}{\left|{\tilde{N}}\right|}
\sum_{i\in{\mathcal{C}}} \left(1-\frac{\mathbf{x}_i^{\text{T}}\cdot\mathbf{z}_i}{\|\mathbf{x}_i\|\|\mathbf{z}_i\|}\right) ^{\gamma}, {\gamma}\geq1,
\end{equation}

\noindent where $\mathbf{x}_i$ denotes the original feature, $\mathbf{z}_i$ is the output of Decoder 1, $\mathcal{C}$ is the set of indices of the masked nodes $\tilde{N}$.

\textbf{Pseudo force recovery.}
The pseudo force is defined as the negative gradient the pseudo energy, which is obtained from the masked node features. We utilize the L1 loss as the criterion for reconstructing the pseudo force:

\begin{equation}
\mathcal{L}_{\text{mf}} = \frac{1}{\left|{\tilde{N}}\right|}\sum_{i\in{C}}
\left\|\hat{\mathbf{f}}_{i}-\mathbf{f}_{i}\right\|,
\end{equation}

\noindent where $\hat{\mathbf{f}}_i$ represents the pseudo forces predicted by Decoder 2 and $\mathbf{f}_i$ are the pseudo forces calculated from the pseudo energy.

\section*{Data availability}
The data other than publicly available data used for experiments in this paper are available at https://cloud.tsinghua.edu.cn/d/7d2bfe81ed3b4269a692/. 

\section*{Code availability}
The source code for reproducing the findings in this paper is available at xxx.

\section*{Acknowledgments}
This work was supported by the National Key R\&D Program of China (NO.2022ZD0160101). M.S. was partially supported by Shanghai Committee of Science and Technology, China (Grant No. 23QD1400900). T.C. and C.T. did this work during their internship at Shanghai Artificial Intelligence Laboratory. 

\section*{Author contributions}
M.S. and S.Z. conceived the idea and led the research. T.C. developed the codes and trained the models. C.T. performed experiments and analyses. Y.L. and X.G. contributed technical ideas for datasets and experiments. D.Z., and W.O. contributed technical ideas for self-supervised methods. T.C., C.T., D.Z., M.S., and S.Z. wrote the paper. All authors discussed the results and reviewed the manuscript. 

\bibliography{sn-bibliography}% common bib file

\end{document}